\documentstyle[12pt,epsfig,axodraw]{article}
\newcommand{\no}     {\nonumber\\}
\newcommand{\ep}	{\epsilon}

\newcommand{\be}{\begin{equation}}
\newcommand{\ee}{\end{equation}}
\newcommand{\bea}{\begin{eqnarray}}
\newcommand{\eea}{\end{eqnarray}}
\newcommand{\Cdot}{\hspace{-1mm}\cdot\hspace{-1mm}}

\newcommand{\jpsi}{{J/\psi}}

\newcommand{\pol}{\rho_{\lambda\lambda^\prime}}

\newcommand{\la}{\lambda}
\newcommand{\sig}{\sigma}
\newcommand{\pr}{\prime}

\begin{document}
\hbox to\hsize{\hfill SNUTP  02-017}
\hbox to\hsize{\hfill DESY 02-071}
\hbox to\hsize{ \hfill hep-ph/0206049}
\hbox to\hsize{\hfill  May  2002}

{\begin{center}
{\large \bf
Joint Angular Distribution of 
$B_s \to J/\psi \phi$ with Subsequent 
$\phi \to \rho \pi$ and
$J/\psi \to \ell^+ \ell^-$ Decays}
\vskip 1cm

G. Kramer \\
{\small \it II. Institut f\"{u}r Theoretische Physik, Universit\"{a}t Hamburg,
D-22761 Hamburg, Germany} \\
\vskip 3mm
H. S. Song \\
{\small \it Center for Theoretical Physics and School of Physics,\\
Seoul National University, Seoul 151-742, Korea} \\
\vskip 3mm
Chaehyun Yu \\
{\small \it Department of Physics, Yonsei University, Seoul 120-749, Korea} \\

\end{center}
\vspace{1cm}

\begin{quote}
The covariant density matrix formalism is used to obtain the joint angular 
distribution of the decay $B_s \to J/\psi \phi(1020)$ with subsequent 
$\phi \to \rho \pi$ and $J/\psi \to \ell^+ \ell^-$ decays. 
The result is shown algebraically to be 
a special case of our previous work on the decay distribution of
$B_d \to J/\psi K_1(1270)$ with $ K_1 \to K \pi $ and  $J/\psi \to \ell^+ 
\ell^-$. 
\end{quote}
}
\vskip 5mm

In the upgraded Tevatron Run II experiment, substantial numbers of 
$B_s \to J/\psi \phi(1020)$ events are expected in the future.  
Previous experimental works have been reported by the CDF Collaboration 
\cite{k1}, and  is well known that the decay process is described by 
three Lorentz invariant terms, {\it i.e.}, 
two CP--even terms and one CP--odd term.  
The joint angular distribution of $B_s \to J/\psi \phi$ with subsequent 
$\phi \to K^+ K^-$ and $J/\psi \to \ell^+ \ell^-$ decays can be described 
\cite{k2} in the same way as that of $B_d \to J/\psi K^\ast$ with subsequent 
$K^\ast \to K\pi$ and $J/\psi \to \ell^+ \ell^-$ decays \cite{k3}, 
and a recent experimental investigation of the latter case has been reported 
by the  Barbar \cite{k4} and the Belle \cite{k5} groups.  
Therefore, the joint angular distribution of $B_s \to  J/\psi \phi$ 
will be determined with high precision in the future.

In the $B_s \to J/\psi \phi$ case, one expects 
$\phi \to \rho \pi$ and $\phi \to \pi \pi \pi$ in addition to 
the dominant $\phi \to K^+ K^-$ decay.  
It will be difficult, if not impossible,
 to identify the $\phi \to \rho \pi$ events 
experimentally at Tevatron even though several hundred 
events are produced in the future. 
However, there is some possibility to construct a $B_s$ factory
in the future where the $\phi \to \rho \pi$ decay can be investigated
more cleanly, and that would give additional information 
on the $B_s \to  J/\psi \phi$ process.

Recently, the covariant density matrix formalism was used to obtain
the joint angular distributions of $B_d \to J/\psi K^\ast(890)$,
as well as $B_d \to J/\psi K_1$(1270),
followed by  their subsequent decays \cite{k6}. 
The purpose of this brief report is to use this method to obtain
the joint angular distribution of $B_s \to J/\psi \phi$ followed
by $\phi \to \rho \pi$ and $J/\psi \to \ell^+ \ell^-$ in the transversity basis
\cite{k2, k3}. 
Since $B_d \to J/\psi K^\ast (890)$ and $B_s \to J/\psi \phi(1020)$ can be 
treated in the same way  for $K^\ast \to K \pi$ and $\phi \to K^+ K^-$,
most of the results in Ref.~6 can be used to obtain the density matrix 
of $J/\psi$ after the $\phi \to \rho \pi$ decay in the process
$B_s \to J/\psi \phi(1020)$. 

As shown in Ref.~6, the angular distribution of the decay $J/\psi \to
\ell^+\ell^-$ is described in terms of the density matrix $\pol$ 
in the transversity basis, defined as in Fig. 1
in the $\jpsi$ rest frame, as follows:
\bea
d {\mit\Gamma} &\sim&  \Big[~ 1+\rho_{00} + (1-3\rho_{00}) \sin^2 \theta_{tr}
\cos^2\phi_{tr} \no
&&\hspace{0.65cm}+2 {\rm Re} (\rho_{1 \mbox{-}1}) 
(\sin^2\theta_{tr} \sin^2\phi_{tr}
-\cos^2 \theta_{tr})\no
&&\hspace{0.65cm}-2 {\rm Im} (\rho_{1 \mbox{-}1}) 
\sin 2\theta_{tr} \sin \phi_{tr}\no
&&\hspace{0.65cm}+\sqrt{2} {\rm Re} (\rho_{1 0}- \rho_{\mbox{-}1 0}) 
\sin^2\theta_{tr} \sin 2\phi_{tr}\no
&&\hspace{0.65cm}-\sqrt{2} {\rm Im} (\rho_{1 0}+ \rho_{\mbox{-}1 0}) 
\sin 2\theta_{tr} \cos \phi_{tr}~\Big],
\label{cccc}
\eea
where $\theta_{tr}$ and $\phi_{tr}$ are the polar and azimuthal angles of the
outgoing lepton $\ell^+$ in the transversity basis as defined in 
Refs.~2 and 3,
where the unit vectors $(\hat{n}_3, \hat{n}_1, \hat{n}_2)$ 
are along the $(\hat{x}, \hat{y}, \hat{z})$
axis. 

The explicit values of the density matrix elements
$\rho_{\la\la^\pr}$ are calculated from the amplitude of the
$\jpsi$ production process. The matrix element of the
$B_s\to \jpsi \phi$ decay is given in terms of
three independent Lorentz scalars,  $A$, $B$ and $C$, as \cite{k7, k8}
\be 
{\cal M} = A \ep_1^\ast\Cdot \ep_2^\ast
+\frac{B}{m_1m_2} \ep_1^\ast \Cdot k \ep_2^\ast \Cdot q
+\frac{iC}{m_1m_2} \ep^{\mu\nu\sig\tau} \ep_{1\mu}^\ast
\ep_{2\nu}^\ast q_\sig k_\tau,
\label{amplitudebjk}
\ee
where $\ep_i (i =1,2)$ are the polarization vectors of $\jpsi$ and $\phi$,
respectively, and $m_1, m_2$ and $q^\mu, k^\mu$ are their masses and momenta. 
Instead of scalar amplitudes $A$, $B$ and $C$
it is more convenient to introduce three transversity amplitudes $A_0,
A_\parallel $ and $A_\perp$ to specify the CP property of the 
decay process. The latter are related to the former by \cite{k2}
\bea
A_0 &=& -xA -(x^2-1)B, \no
A_\parallel &=& \sqrt{2} \hspace{1mm}A, \no
A_\perp &=& \sqrt{2(x^2-1)} \hspace{1mm}C,
\eea
where $x$ is defined as
\be
x= \frac{k \Cdot q}{m_1 m_2} = \frac{ m_B^2 - m_1^2 -m_2^2}{2m_1 m_2}.
\ee
When the decay of the $B_s$ into $\jpsi$ and $\phi$ is specified by the 
polarization vectors $\ep_1(q,\la_1)$ and $\ep_2(k,\la_2)$, respectively,
the joint density matrix elements of  $\jpsi$ and $\phi$  can be
obtained from the square of Eq.~(2) and by using the following relation
\cite{k9} for each $\ep_i ( i= 1,2)$:

\be
\ep^\mu(q,\la)\ep^{\nu\ast}(q,\la^\pr)
=\frac{1}{3}\bigg(-g^{\mu\nu}+\frac{q^\mu q^\nu}{m_1^2}\bigg) 
\delta_{\la^\pr\la}
-\frac{i}{2m_1}\ep^{\mu\nu\sig\tau}q_\sig n_\tau^i (S^i)_{\la^\pr\la}
-\frac{1}{2}n_i^\mu n_j^\nu (S^{ij})_{\la^\pr\la},
\label{projection}
\ee
where the $ (S^i )$ are the standard matrix representations 
of the spin-1 angular momentum operators
and $( S^{ij}) $ are traceless and symmetric matrices. 

Then, one obtains 
\bea
\rho_{\la_1\la_1^\pr, \la_2\la_2^\pr}&\sim&
\frac{1}{9} (|A_0|^2+|A_\parallel|^2+|A_\perp|^2)
\delta_{\la_1\la_1^\pr, \la_2\la_2^\pr} \no
&&-\frac{2}{3\sqrt{\mit\Delta}}{\rm Re} (A_\parallel^\ast A_\perp)
\{ m_1 n_1^i \Cdot k \hspace{1mm}( S^i)_{\la_1\la_1^\pr}
\delta_{\la_2\la_2^\pr} 
+ m_2 n_2^k \Cdot q \delta_{\la_1\la_1^\pr}
(S^k)_{\la_2\la_2^\pr}\} \no
&&+\frac{1}{\mit\Delta} 
\Big\{ \hspace{3mm}(|A_\parallel|^2+|A_\perp|^2)
m_1m_2 \hspace{1mm}n_1^i\Cdot k n_2^k\Cdot q \no
&&\hspace{1cm}+\sqrt{\frac{\mit\Delta}{2}} {\rm Im}(A_0^\ast A_\perp)
\langle q k n_1^i n_2^k\rangle \no
&&\hspace{1cm}-\frac{\sqrt{2}}{4}{\mit\Delta} {\rm Re}(A_0^\ast A_\parallel)
(n_1^i\Cdot n_2^k -\frac{4q\Cdot k}
{\mit\Delta} n_1^i\Cdot k n_2^k \Cdot q)\Big\}
(S^i)_{\la_1\la_1^\pr} (S^k)_{\la_2\la_2^\pr} \no
&&-\frac{1}{3{\mit\Delta}}
(2|A_0|^2-|A_\parallel|^2-|A_\perp|^2)
\Big\{ ~m_1^2\hspace{1mm} n_1^i\Cdot k \hspace{1mm}n_1^j\Cdot k\hspace{1mm} (S^{ij})_{\la_1\la_1^\pr}
\delta_{\la_2\la_2^\pr} \no
&&\hspace{5.35cm}+m_2^2\hspace{1mm} n_2^k\Cdot q\hspace{1mm} n_2^l\Cdot q\hspace{1mm} \delta_{\la_1\la_1^\pr}
(S^{kl})_{\la_2\la_2^\pr} \Big\} \no
&&-\frac{1}{\mit\Delta}  \Big\{
~\sqrt{2} {\rm Im} (A_0^\ast A_\parallel) \langle q k n_1^i n_2^k \rangle \no
&&\hspace{0.9cm} +\frac{2}{\sqrt{\mit\Delta}}
{\rm Re}(A_\parallel^\ast A_\perp) m_1 m_2 n_1^i\Cdot k n_2^k\Cdot q \no
&&\hspace{0.9cm}-\sqrt{\frac{\mit\Delta}{2}}
{\rm Re} (A_0^\ast A_\perp) 
\Big( n_1^i\Cdot n_2^k - \frac{4q\Cdot k}{\mit\Delta}
n_1^i\Cdot k n_2^k \Cdot q \Big) \Big\} \no
&&\hspace{5mm}\times \{ m_2 n_2^l\Cdot q\hspace{1mm} (S^i)_{\la_1 \la_1^\pr}
(S^{kl})_{\la_2\la_2^\pr} 
+ m_1 n_1^j\Cdot q\hspace{1mm} (S^{ij})_{\la_1 \la_1^\pr}
(S^{k})_{\la_2\la_2^\pr} \} \no
&&+\Big\{ \frac{4}{{\mit\Delta}^2}|A^0|^2 m_1^2 m_2^2
n_1^i \Cdot k n_1^j\Cdot k n_2^k \Cdot q n_2^l\Cdot q \no
&&\hspace{3mm} +\frac{1}{8} |A_\parallel|^2
( n_1^i\Cdot n_2^k -\frac{4q\Cdot k}{\mit\Delta}
  n_1^i \Cdot k n_2^k \Cdot q)
( n_1^j\Cdot n_2^l -\frac{4q\Cdot k}{\mit\Delta}
  n_1^j \Cdot k n_2^l \Cdot q) \no
&&\hspace{3mm} +\frac{1}{2{\mit\Delta}} |A_\perp|^2
\langle q k n_1^i n_2^k\rangle \langle q k n_1^j n_2^l\rangle \no
&&\hspace{3mm} -\frac{\sqrt{2}}{2{\mit\Delta}}{\rm Re}(A_0^\ast A_\parallel)
m_1 m_2 n_1^i \Cdot k n_2^k\Cdot q
( n_1^j\Cdot n_2^l -\frac{4q\Cdot k}{\mit\Delta}
  n_1^j \Cdot k n_2^l \Cdot q) \no
&&\hspace{3mm} +\frac{2\sqrt{2}}{{\mit\Delta} \sqrt{\mit\Delta}}
{\rm Im} (A_0^\ast A_\perp)
m_1 m_2 n_1^i \Cdot k n_2^k\Cdot q
\langle q k n_1^j n_2^l\rangle \no
&&\hspace{3mm}-\frac{1}{2 \sqrt{\mit\Delta}}
{\rm Im} (A_\parallel^\ast A_\perp)
( n_1^i\Cdot n_2^k -\frac{4q\Cdot k}{\mit\Delta}
  n_1^i \Cdot k n_2^k \Cdot q)
\langle q k n_1^j n_2^l\rangle  \Big\} 
\no
&&\hspace{3mm} \times
(S^{ij})_{\la_1\la_1^\pr}
(S^{kl})_{\la_2\la_2^\pr}.
\label{dd}
\eea
Here, the density matrix is multiplied  by 
$(|A_0|^2+|A_\parallel|^2+|A_\perp|^2)$ for convenience, and 
${\mit\Delta}$ and $\langle q k n_1^j n_2^l\rangle $ are defined as
\bea
&&{\mit\Delta} = (m_B^4 +m_1^4 +m_2^4 -2m_B^2 m_1^2-2m_B^2 m_2^2-2m_1^2 m_2^2)
\no
&&\hspace{4.6mm}= 4[(q.k)^2 - m_1^2 m_2^2] = 4m_1^2 m_2^2 (x^2-1),
\label{delta} \\
&&\langle q k n_1^j n_2^l\rangle 
= \ep^{\mu\nu\sig\tau} q_\mu k_\nu (n_\sig^j)_1 (n_\tau^l)_2.
\eea

The amplitude of the decay process
$\phi(k,\la_2)\to \rho(\tilde{k}^\pr)\pi(\tilde{k})$ is described by
\be
{\cal M}_2 = i f \ep^{\mu\nu\la\tau} \ep_{2\mu} (k) {\ep_{3\nu}}^* (\tilde{k}^\pr) k_\la  
\tilde{k}_\tau,
\label{ampk1}
\ee
where $\ep_3^\mu$ and $\tilde{k}^\pr$
are the polarization vector and the momentum of $\rho$, respectively.
After the polarization of  $\rho$ is summed over, one obtains
\be
\rho_{\la_2^\pr \la_2}^D (\phi)= \frac {1}{3} \delta_{\la_2^\pr\la_2} 
+ \frac {m_{\phi}^2 }{\bar{\Delta}} \tilde{k} \Cdot n_i \tilde{k} \Cdot n_j
(S^{ij})_{\la_2^\pr\la_2},
\ee
where $\bar{\mit\Delta}$ is defined as 
\bea 
{\bar{\mit\Delta}} = {m_{\phi}^4 +m_{\rho}^4 +m_{\pi}^4 
-2 m_{\phi}^2 -2m_{\rho}^2 -2 m_{\pi}^2 }.
\eea 
The result of Eq.~(10) can be obtained from Eq.~(6).
Then, the density matrix of $\jpsi$ can be obtained from the product
\be
\rho_{\la_1 \la_1^\pr, \la_2 \la_2^\pr}
\rho_{\la_2^\pr \la_2}^D(\phi).
\ee
The joint angular distribution for the decay process
$B_s\to \jpsi\phi~ (\jpsi \to \ell^+ \ell^-, \phi \to \rho \pi)$
in the transversity basis becomes
\bea
\frac{1}{\mit\Gamma}\frac{d {\mit\Gamma}}
{d \cos \theta_1 d \cos \theta_{tr} d \phi_{tr}}
&=&
\frac{9}{64\pi}\bigg[~ 
2|A_0|^2 \bigg\{ \sin^2 \theta_1
(1-\sin^2\theta_{tr} \cos^2\phi_{tr})\bigg\} \nonumber \\
&&\hspace{8.25mm}
+~|A_\parallel|^2\bigg\{\cos^2\theta_1(1-\sin^2\theta_{tr}
\sin^2\phi_{tr}) +\sin^2 \theta_{tr} \bigg\} \no
&&\hspace{8.25mm}+~|A_\perp|^2\bigg\{1-\sin^2\theta_{tr} 
\sin^2\phi_{tr} +\cos^2 \theta_1 \sin^2 \theta_{tr} 
\bigg\} \no
&&\hspace{8.25mm}-~{\rm Im}\Big(A_\parallel^\ast A_\perp\Big)
\sin^2\theta_1 \sin 2\theta_{tr} \sin \phi_{tr} \nonumber \\
&&\hspace{8.25mm}+~\frac{1}{\sqrt{2}}{\rm Re}\Big(A_0^\ast A_\parallel\Big)
\sin 2\theta_1\sin^2\theta_{tr} \sin 2\phi_{tr} \nonumber \\
&&\hspace{8.25mm}-~\frac{1}{\sqrt{2}}{\rm Im}\Big(A_0^\ast A_\perp\Big)
\sin 2\theta_1\sin 2\theta_{tr} \cos \phi_{tr} ~\bigg],
\label{distribution}
\eea
where $\theta_1$ is the angle between the $\phi$ direction and the $\pi$ direction
in the $\phi$ rest frame, and the three amplitudes are normalized as
$|A_0|^2+|A_\parallel|^2+|A_\perp|^2 =1$.
It is noted that in Ref.~6 we have considered the process
$B\to \jpsi K_1(1270)$ and $\jpsi\to \ell^+ \ell^-$,
$K_1\to K \pi$ and Eq.~(13) is a special case obtained  by putting $\alpha
=  -1$ in Eq.~(28) or $\xi$ = {3/2} in Eq.~(29) of Ref.~6.
This method can be used to discuss the polarization effects in $W$ boson
productions at $e^+ e^- $ linear collider \cite{k10}.

\section*{Acknowledgments}
The authors would like to thank Prof. Intae Yu for discussion related         
with the CDF experiments.  The work is supported in part by the DFG-KOSEF     
Collaboration (2000) under contracts 446 KOR-113/137/0-1 (DFG) and        
20005-111-02-2 (KOSEF). HS was supported in part  by  the Korea  Research  
Foundation (KRF-2000-D00077) and by the BK21 program.   
The work of CY was supported by grant No. R02-2002-000-00168-0 of the KRF. 
The authors thank their guest institutes for the warm hospitality during      
their visits.

\newpage
{\Large \bf Figure Captions}
\vskip 2cm

\begin{description}

\item
Fig. 1.
Definitions of the angles in the transversity basis.
$\theta_{tr}$ and $\phi_{tr}$ are the polar and 
the azimuthal angles of $\ell^+$ in the $\jpsi$ rest frame.
$\theta_1$ is the angle between the $\phi$ direction and the $\pi$ direction
in the $\phi$ rest frame.

\end{description}

\begin{center}
\begin{figure}[htb]
\vspace{1cm}
\psfig{file=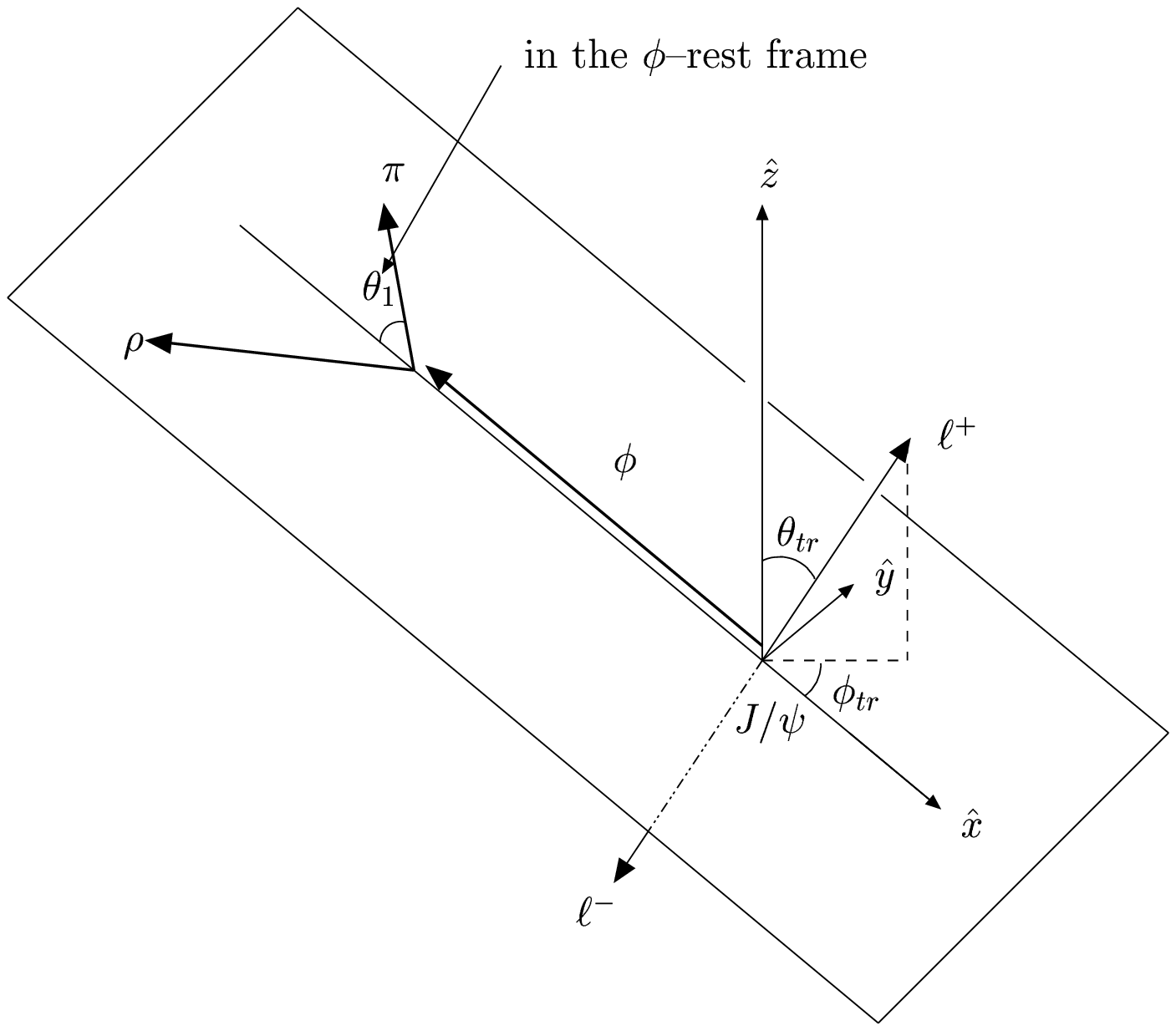}
\vspace{0.5cm}
\caption{\it
}
\label{fig1}
\end{figure}
\end{center}

\end{document}